\documentclass[useAMS,usenatbib]{mn2e}
\usepackage{url}
\usepackage{graphicx}
\usepackage{verbatim}
\usepackage[usenames]{color}
\usepackage[flushleft]{threeparttable}
\DeclareGraphicsExtensions{.pdf,.png,.jpg,.mps,.eps,.ps}
\usepackage{amsmath}
\usepackage{natbib}
\usepackage{color}
\usepackage{graphicx}
\usepackage{lscape}
\usepackage{gensymb}
\usepackage{amssymb}

\newcommand\nicer{{\it NICER}}
\newcommand\nustar{{\it NuSTAR}}
\newcommand\swift{{\it SWIFT}}

\newcommand\kev{{\rm~keV}}

\newcommand\kms{\ifmmode {\rm~km\ s}^{-1} \else ~km s$^{-1}$\fi}
\newcommand\Hunit{\ifmmode {\rm~km\ s}^{-1}\ {\rm Mpc}^{-1}
        \else ~km s$^{-1}$ Mpc$^{-1}$\fi}
\newcommand\ctssec{\ifmmode {\rm~count\ s}^{-1} \else ~count s$^{-1}$\fi}
\newcommand\ergsec{\ifmmode {\rm~erg\ s}^{-1} \else
        ~erg s$^{-1}$\fi}
\newcommand\funit{\ifmmode {\rm~erg\ s}^{-1}\;{\rm cm}^{-2} \else
        ~ergs s$^{-1}$ cm$^{-2}$\fi}
\newcommand\phflux{\ifmmode {\rm~photon\ s}^{-1}\;{\rm cm}^{-2}
        \else   ~photon s$^{-1}$ cm$^{-2}$\fi}
\newcommand\efluxA{\ifmmode {\rm~erg\ s}^{-1}\;{\rm cm}^{-2}\;{\rm
        \AA}^{-1} \else ~erg s$^{-1}$ cm$^{-2}$ \AA$^{-1}$\fi}
\newcommand\efluxHz{\ifmmode {\rm~erg\ s}^{-1}\;{\rm cm}^{-2}\;{\rm
        Hz}^{-1} \else ~erg s$^{-1}$ cm$^{-2}$ Hz$^{-1}$\fi}
\newcommand\cc{\ifmmode {\rm~cm}^{-3} \else cm$^{-3}$\fi}
\newcommand\FWHM{\ifmmode {\rm~FWHM} \else ${\rm~FWHM}$\fi}
\newcommand\Msun{\ifmmode M_{\odot} \else $M_{\odot}$\fi}
\newcommand\Lsun{\ifmmode L_{\odot} \else $L_{\odot}$\fi}

\newcommand\hbeta{\ifmmode {\rm H}\beta \else H$\beta$\fi}
\newcommand\Kalpha{\ifmmode {\rm K}\alpha \else K$\alpha$\fi}
\newcommand\nh{\ifmmode N_{\rm H} \else N$_{\rm H}$\fi}

\usepackage{graphicx}
\usepackage{url}
\usepackage{verbatim}
\usepackage{color}

\title[Complex spectral behavior of Swift~J1858.6-0814]{The complex spectral behavior of the newly discovered neutron star X-ray binary Swift~J1858.6-0814}
\author[Mondal et al.]{\parbox[]{6.5in}{Aditya S. Mondal$^{1}\thanks{E-mail: adityas.mondal@visva-bharati.ac.in}$, B. Raychaudhuri$^{1}$, G. C. Dewangan$^{2}$   \\
\small
$^{1}$Department of physics, Visva-Bharati, Santiniketan, West Bengal-731235, India \\
$^{2}$Inter-University Centre for  Astronomy \& Astrophysics (IUCAA), Pune, 411007 India \\
}}

\date{\today}
\begin{document}
\maketitle
\begin{abstract}
We report on the \nustar{} observation of the newly discovered neutron star X-ray binary Swift~J1858.6-0814 taken on 23rd March 2019. The light curve of the source exhibits several large flares during some time intervals of this observation. The source is softer in the high-intensity interval where the large flaring activity mainly occurs. We perform time-resolved spectroscopy on the source by extracting spectra for two different intensity intervals. The source was observed with a $3-79 \kev{}$ luminosity of $\sim 9.68\times 10^{36}$ ergs/s and $\sim 4.78\times 10^{36}$ ergs/s for high and low-intensity interval, respectively assuming a distance of $15$ kpc. We find a large value of the absorbing column density ($\rm{N_{H}}\sim 1.1\times 10^{23}$ cm$^{-2}$), and it appears to be uncorrelated with the observed flux of the source. Each spectrum shows evidence of Fe K$\alpha$ emission in the $5-7$\kev{} energy band, an absorption edge around $\sim 7-8$\kev{}, and a broad Compton hump above $15$\kev{}, indicating the presence of a reflection spectrum. The observed features are well explained by the contribution of a relativistic reflection model and a partially covering absorption model. From the best-fit spectral model, we found an inner disc radius to be $4.87_{-0.96}^{+1.63}\;R_{ISCO}$ (for the high-intensity interval) and $5.68_{-2.78}^{+9.54}\;R_{ISCO}$ (for the low-intensity interval), indicating a significant disc truncation. The disk inclination is found to be relatively low, $i< 33^{0}$. We further place an upper limit on this source's magnetic field strength considering the disc is truncated at the magnetospheric radius.
\end{abstract}

\begin{keywords}
  accretion, accretion discs - stars: neutron - X-rays: binaries - stars:
  individual Swift~J1858.6-0814
\end{keywords}
\section{Introduction}
Low-mass X-ray binaries (LMXBs) are systems that contain a neutron star~(NS) or a black hole~(BH) accreting from a low-mass ($\lesssim 1 \Msun$) companion star via Roche-lobe overflow. Many observable properties are similar in either case. Thus, the determination of compact objects in binary systems is often a challenging task. However, accreting systems have many unique properties which can give a definitive determination of whether the accreting object is a BH or an NS. The unique features related to those hosting NSs are the following: (i) The surface of the NS can provide an additional location for emission component, and they can support large-scale magnetic fields \citep{2009ApJ...694L..21C, 2009A&A...493L..39P, 2017ApJ...847..135L}. (ii) NSs can pulse coherently on their spin period, which can be observed at wavelengths from radio to X-ray \citep{1968Natur.217..709H, 1998Natur.394..344W, 1996ApJ...469L...9S}. (iii) Detection of Type-I X-ray bursts due to explosive thermonuclear burning of the accreted material on the NS surface \citep{1976ApJ...205L.131G, 2006csxs.book..113S, 2008ApJS..179..360G}. \\

LMXBs may be persistent accreator or transient systems. Most BH and some NS LMXBs are transient systems. X-ray transients spend most of their lives in quiescence but show bright outbursts lasting for weeks to months. Any instabilities in the disc cause an increase in the accretion rate, causing the system to go into an outburst. Persistent accretor may have an X-ray luminosity of $L_{X}\gtrsim 10^{36} \ergsec{}$ (\citealt{2019ApJ...873...99L, 2017ApJ...836..140L}). In contrast, transient systems often undergo cycles of outburst and quiescence due to the modulation in the rate at which matter from the companion star accretes onto the compact object (either NS or BH). Transient LMXBs undergo recurrent bright ($L_{X}\gtrsim 10^{36} \ergsec{}$) outbursts lasting from days to weeks and then return to long intervals of X-ray quiescence ($L_{X}\lesssim 10^{34} \ergsec{}$) lasting from months to years \citep{2010A&A...524A..69D}. The long-term average mass accretion rate of the  transient systems is significantly lower than that in the persistent systems. \\

Swift J1858.6-0814 is a Galactic transient LMXB \citep{2019ATel12704....1F, 2018ATel12167....1B, 2020ATel13725....1P}. It has many interesting properties, such as erratic flaring on the minute time scale, evidence for wind outflows, strong and variable local absorption, and an anomalously hard X-ray spectrum \citep{2018ATel12164....1V, 2018ATel12184....1B, 2018ATel12158....1L, 2018ATel12220....1R, 2020ApJ...890...57H, 2020ApJ...893L..19M, 2023MNRAS.520..542S}. This source was initially discovered as an X-ray transient in October 2018 \citep{2018ATel12151....1K} with the Burst Alert Telescope (BAT) aboard the Neil Gehrels \swift{} observatory \citep{2004ApJ...611.1005G}. The source was identified with a variable counterpart observed at optical and radio wavelengths \citep{2018ATel12164....1V, 2018ATel12186....1R, 2018ATel12184....1B}. Initially, the source was described as an analogue of the BH sources V4641 Sgr and V404 Cyg because of the similarity observed in strong X-ray variability during their outbursts \citep{2000ApJ...528L..93W, 2016MNRAS.459..554G}. Moreover, the large amplitude flaring in the light curve, hard spectrum, and prominent reflection features resemble the behavior of the BH sources V4641 Sgr and V404 Cyg. The source was initially notable for its unusually strong X-ray variability. The X-ray spectra of Swift J1858.6-0814 were hard: $\Gamma\sim (1-1.2)$ and showed strong, variable obscuration. They showed evidence of strong disc reflection and local absorption features: Iron K$\alpha$ emission line, K edge, and soft X-ray emission lines \citep{2018ATel12220....1R, 2020ApJ...890...57H, 2020MNRAS.498...68B}. Detection of significant intrinsic local absorption in both the optical and X-ray spectrum suggested that a significant amount of mass was ejected from the inner accretion flow \citep{2020ApJ...893L..19M, 2020ApJ...890...57H}.\\

Swift J1858.6-0814 remained active up to August 2019 as observed by \nustar{} and then entered the Sun constraint for most X-ray telescopes in November 2019. The Monitor of the All-sky X-ray Imager (MAXI) detected the source again in February 2020 with a significantly less variable state. So, the source was then transitioned from the so-called flaring outburst state (2018-2019) to a steady outburst state (2020) with a more persistent X-ray flux \citep{2020ATel13455....1N, 2020MNRAS.498...68B}. Before 2020, the source was only known for atypically large variability. In March 2020, several Type-I X-ray bursts were detected during the steady state with the \nicer{} and the \nustar{}, and thus, the source was identified as an NS LMXB \citep{2020MNRAS.499..793B}. These bursts exhibit photospheric radius expansion, allowing a distance estimate of $\sim 15$ kpc \citep{2020MNRAS.499..793B}. Strong periodic drops in X-ray flux were also detected, consistent with eclipses by the secondary star and variable obscuration due to the thickness of the disc or the accretion stream \citep{2021MNRAS.503.5600B}. \citet{2021MNRAS.503.5600B} also suggest that the variable obscuration is responsible for the unusually strong variability in Swift J1858.6-0814. The source is highly variable at radio wavelengths in most observations. Radio emission showed variability up to a factor of 8 within 20 minutes and is consistent with a compact jet \citep{2020MNRAS.496.4127V}. Optical emission showed fast flares (stronger at longer wavelengths) and variability, which changes by factors of several within minutes \citep{2018ATel12197....1P}. The source also exhibited wind-formed, blue-shifted absorption lines associated with C IV, N V, and He II in time-resolved UV spectroscopy during a luminous hard state, which is interpreted as a warm, moderately ionized outflow component in this state \citep{2022Natur.603...52C}. \\

Previously, \citet{2020ApJ...890...57H} performed the spectral analysis from the flaring and the non-flaring states based on the first NuSTAR observation, considering BH as the compact object for this source. The identification of Swift~J1858.6-0814 as an NS X-ray binary means that the spectral behaviour of this source must be explained in a model which is compatible with an NS accretor. The spectral behavior and the underlying mechanisms may be understood from the processes related to NS accretor as some differences exist between accreting BH and NS systems. In a BH system, the accretion disc is expected to be sharply truncated at the innermost stable circular orbit (ISCO). In contrast, in an NS system, the disc's inner edge can reach the star's surface if not impeded by a boundary layer or the magnetosphere \citep{2001ApJ...547..355P, 2013A&A...550A...5E, 2009ApJ...694L..21C, 2009MNRAS.400..492I}. Moreover, NSs have much lower spin values than black holes \citep{2008ApJS..179..360G, 2011ApJ...731L...7M}. Due to this slower rotation speed and the solid surface of the NS, the inner edge of the disc is at a much higher radius than that for a rapidly rotating BH. \\

In this work, we present the spectral analysis of this source using \nustar{} observation, considering these relevant issues in the context of X-ray reflection. We try to put constraints on the physical parameters (e.g., the inner radius of the accretion disc, the ionization of the plasma in the disc, and the inclination of the system) based on relativistic reflection modeling. Moreover, the magnetic field strength in Swift J1858.6-0814 has not yet been measured. Here we measure it from the reflection modeling. We comment on the characterization of the complex spectral states because of this source's large-scale flaring behavior. This paper is organized as follows: The observation and data reduction are discussed in Section 2. Section 3 and Section 4 are focused on the details of the timing and spectral analysis, respectively. In section 5, we discuss the results obtained from the analysis. \\ 

\begin{figure*}
\centering
\includegraphics[scale=0.65, angle=0]{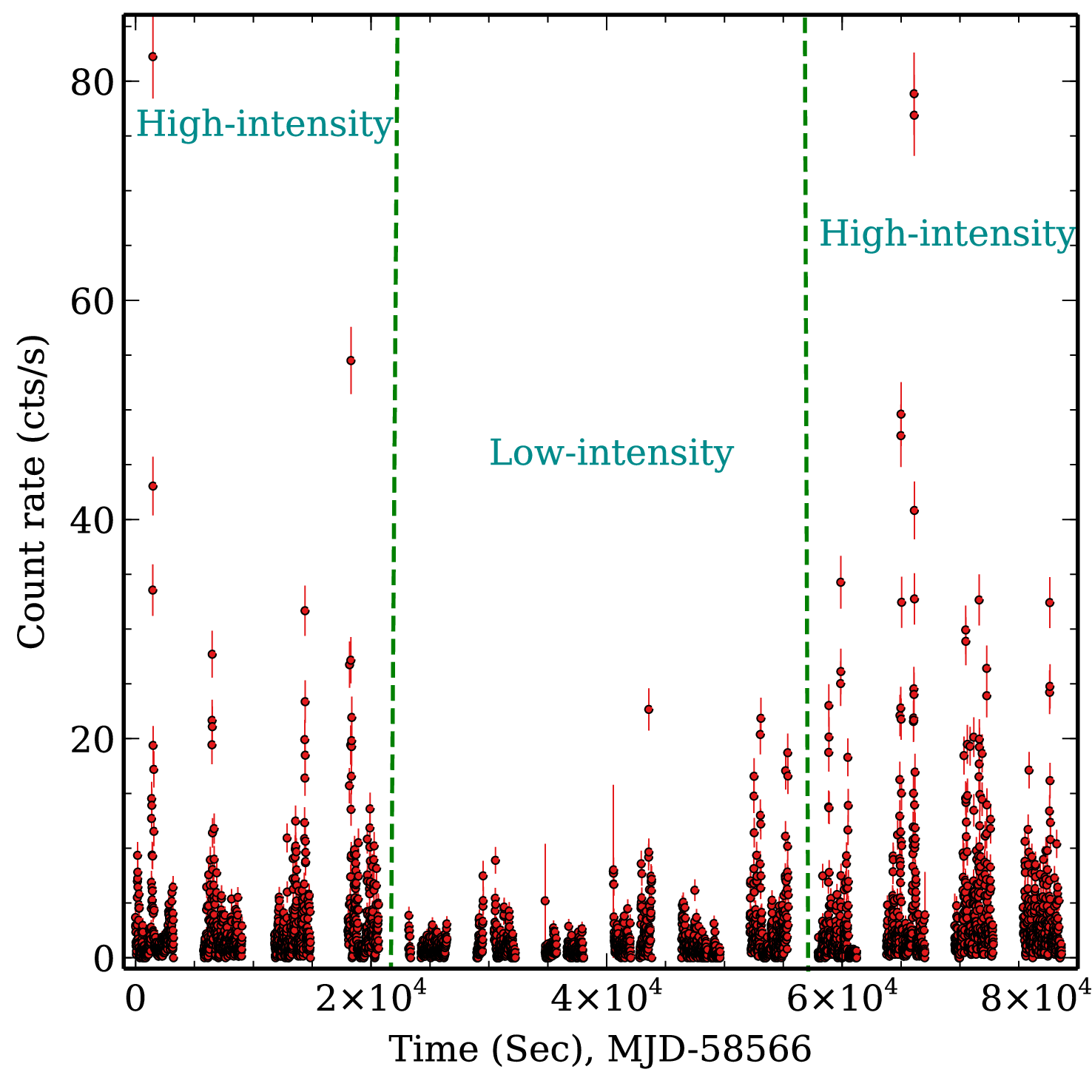}
\caption{$3-79\kev{}$ \nustar{}/FPMA light curve of the source with 10s time bins. The light curve reveals that the source is variable and shows large amplitude flares during this observation. Dashed lines indicate two different intensity intervals from which energy spectra have been extracted. } 
\label{Fig1}
\end{figure*}

\section{observation and data reduction}
The Galactic X-ray transient Swift~J1858.6-0814 was observed with the \textit{Nuclear Spectroscopic Telescope ARray} (\nustar{}; \citealt{2013ApJ...770..103H}) for a number of times in between the years 2018 to 2020. It was first observed by \nustar{} on 3rd November 2018 for $\sim 52$ ks. This coordinated observation (obsID: 80401317002) was analyzed and presented by \citet{2020ApJ...890...57H}. Later time \nustar{} observed the source five times in the year 2019. We have chosen one of these observations carried out on 23rd March 2019 (obs ID: 90501309002) for our analysis. In this particular observation, the source was observed by \nustar{} for a total exposure of $\sim 37$ ks. \\

The \nustar{} data were collected using two hard X-ray imaging telescopes on board \nustar{}, i.e., the focal plane mirrors (FPM) A and B. We reduced the data using the standard data analysis software {\tt NUSTARDAS v2.1.1} task (included in {\tt HEASOFT v6.29}) and using the latest {\tt CALDB} version available. The calibrated and screened event files have been generated using the task {\tt nupipeline}. The source was selected from a circular area of $100''$ radius for both module FPMA and FPMB, centered at the source coordinates. To account for the background, we used a circular area of the same radius but far away from the source position (to ensure negligible contamination from the source) for both instruments. We then used the tool {\tt nuproducts} to build the filtered event files, the background subtracted light curves, the spectra, and the arf and rmf files. To extract spectra from the high and low intensity intervals, we applied a Good Time Interval (GTI) file, which was created by FTOOLS {\tt maketime}. We grouped the FPMA and FPMB spectral data with a minimum of $50$ counts per bin. Finally, the FPMA and FPMB spectra are fitted simultaneously, leaving a floating cross-normalization constant.

\section{Temporal Analysis}
\begin{figure}
\centering
\includegraphics[scale=0.30, angle=-90]{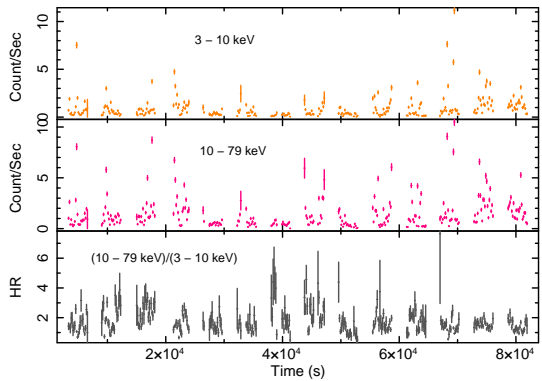}
\caption{This figure's top and the middle panel shows the source's count rate with time in the $3-10$ \kev{} and $10-79$ \kev{} energy band, respectively. The bottom panel shows the variation of the hardness ratio which is defined here as the $10-79$ \kev{} count rate divided by the $3-10$ \kev{} count rate with time.} 
\label{Fig2}
\end{figure}

\begin{figure}
\centering
\includegraphics[scale=0.40, angle=0]{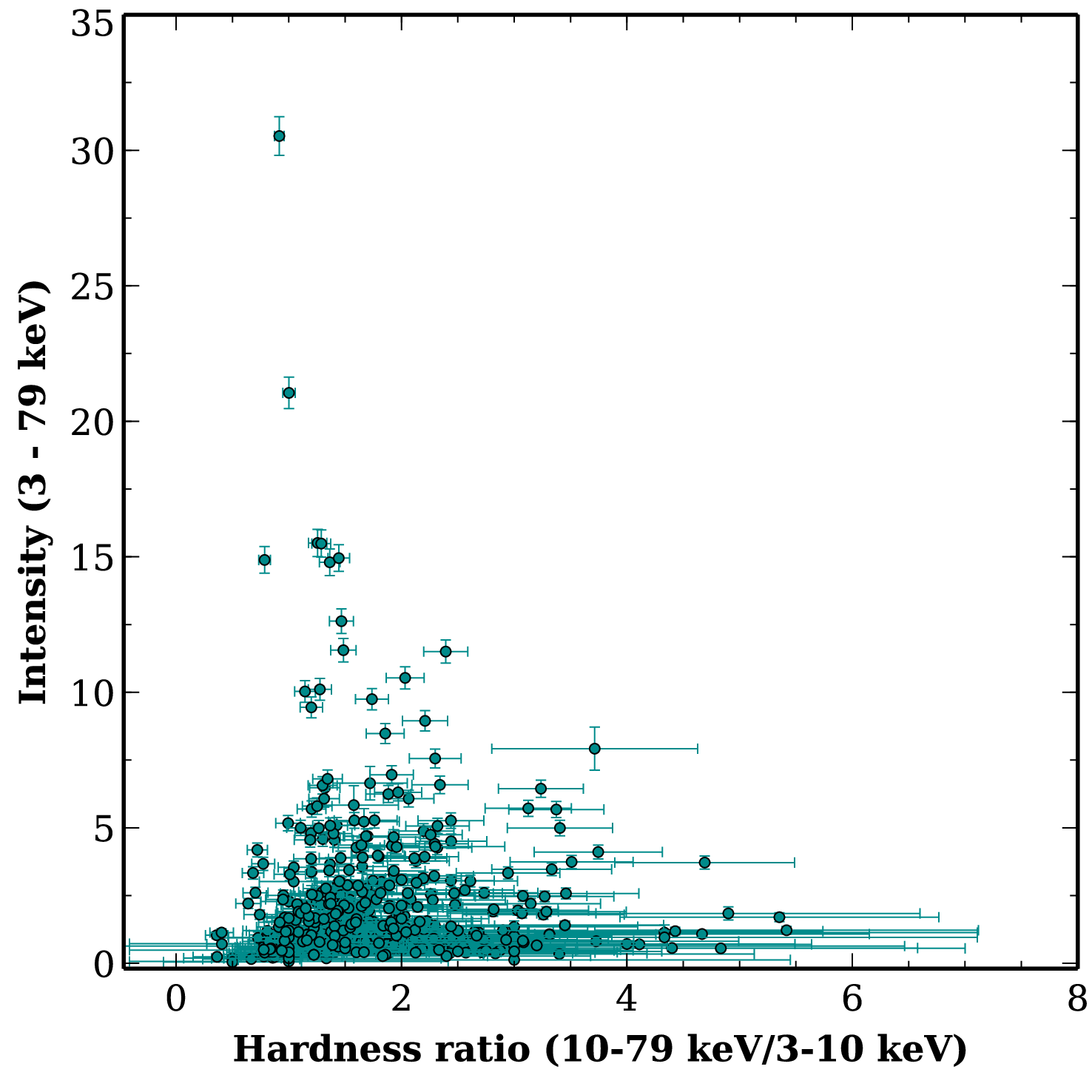}
\caption{Hardness Intensity diagram: shows the variation of $3-79$ \kev{} energy band count rate with the hardness ratio. } 
\label{Fig3}
\end{figure}

Figure~\ref{Fig1} shows the long-term 10s bin-size \nustar{} light curve of the source in the 3-79 \kev{} energy band. The light curve shows prominent flaring on time scales of seconds. One of the flares exhibited an increase in count rate by a factor of $\sim 40$ to the source's average count rate ($\sim 2$ cts/s). This observation samples the source in two states (as the dashed line in Figure~\ref{Fig1} indicates): A high-intensity interval where the source exhibits flaring above $25$ \ctssec{} and a low-intensity interval with a count rate below $25$ \ctssec{} during flaring. After being split, the high and the low intensity intervals consisted of exposure times of $\sim 19.3$ ks and $\sim 16.7$ ks, respectively. The highly flaring activity mainly occurs in the high-intensity interval, while the low-intensity interval shows less flaring. We also extracted the \nustar{} 100s bin-size light curves in the $3-10$ \kev{} and $10-79$ \kev{} energy bands and calculated the hardness ratio (HR). As apparent from the HR, the source showed small changes in its hardness ratio during this observation, but no clear trend with time was observed (see Figure~\ref{Fig2}). In particular, the HR is comparatively larger for the low-intensity interval with less flaring activity. In addition, we plotted the 3-79 \kev{} flux versus the HR to produce the hardness-intensity diagram (HID). We found from the HID (see Figure~\ref{Fig3}) that the hardness ranges from $\sim 0.6$ at the highest flaring stage (associated with the high-intensity interval) to $\sim 6$ at its final stage (associated with the low-intensity interval). We also note that the source is softer in the high-intensity interval where the flaring activity mainly occurs. In other words, we can say that the bright flares lie in the high-intensity interval and are always relatively soft. We thus conduct a time-average spectral analysis for two different intensity intervals with this \nustar{} observation.

\section{spectral analysis}
To fit the \nustar{} FPMA and FPMB spectra simultaneously between $3.5$ to $60$ \kev{} energy band, we used the spectral analysis package {\tt XSPEC} $v12.12.0$ \citep{1996ASPC..101...17A}. To coordinate the calibration differences in different instruments, we used a model {\tt constant}, which we fixed to $1$ for the \nustar{} FPMA and allowed it to vary for the \nustar{} FPMB. We modelled the Galactic interstellar medium absorption by the model {\tt TBabs} with {\tt wilm} abundances \citep{2000ApJ...542..914W} and {\tt vern} \citep{1996ApJ...465..487V} cross section. Spectral uncertainties are given at $90$ percent confidence intervals unless otherwise stated.\\

\subsection{Continuum modeling}
\begin{figure*}
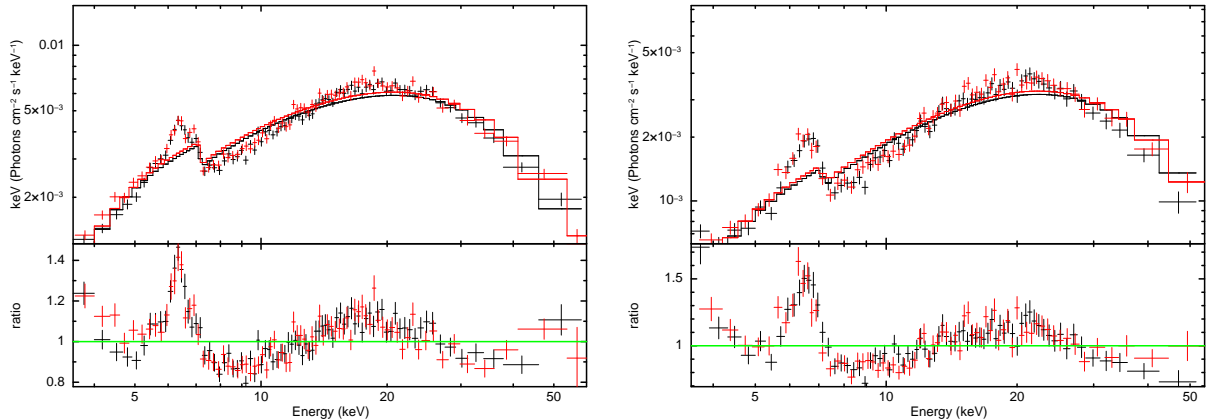

\includegraphics[scale=0.32, angle=-90]{fig4.ps}
\includegraphics[scale=0.32, angle=-90]{fig5.ps}
\caption{The spectrum obtained from the high-intensity and the low-intensity interval is presented in the left and right panels, respectively. Continuum is fitted with the thermal Comptonization model {\tt nthcomp}, which describes the high-energy shapes and the low-energy rollover more accurately compared to a {\tt powerlaw} or {\tt cutoffpl} model. Fit ratios associated with this continuum model are shown in the bottom panel of each plot. The presence of reflection features is evident in both spectra. The data were rebinned for better visualisation.} 
\label{Fig4}
\end{figure*}

We extracted the time-averaged energy spectra for two different intensity intervals. The spectra extracted from the high-intensity interval ($> 25$ \ctssec{} during flares) and the low-intensity interval ($<25 \ctssec{}$ during flares) had exposure times of $\sim 19.3$ ks and $\sim 16.7$ ks, respectively, as stated earlier. A variety of models have been previously used to describe the continuum emission. The hard/Comptonized emission was previously modeled mainly by the power-law component. Here, to probe the spectral shape of the source in a better way, we start our analysis by fitting both the spectra with the thermal Comptonization model {\tt nthcomp} (\citealt{1996MNRAS.283..193Z, 1999MNRAS.309..561Z}). This model describes the high-energy shapes and the low-energy rollover more accurately compared to a power-law or an exponentially cut-off power-law model. Moreover, this model allows us to select the spectral shape of the source of seed photons between a {\tt BBODY} and a {\tt DISKBB}. During the fitting process, we used the {\tt DISKBB} thermal component as a source of seed photons for {\tt nthcomp}. The emission from the NS surface/boundary layer is usually modeled by the {\tt BBODY} component, which becomes significant for the soft spectral state. For this observation, the source is in a hard spectral state that precludes illumination by the NS surface/boundary layer. This model well describes the continuum with $\chi^2/dof=1374/580$ and $\chi^2/dof=876/283$ for the spectrum with high and low intensity intervals, respectively. The {\tt nthcomp} model yields power law photon index $\Gamma=1.3\pm 0.02$, electron temperature of the corona $kT_{e}=10.5\pm 0.16$\kev{}, and a seed photon temperature of the Comptonization $kT_{bb} < 0.18$\kev{} for the high-intensity interval. The corresponding values for the low-intensity interval are $\Gamma=1.2\pm0.01$, $kT_{e}=9.8^{+0.14}_{-0.18}$\kev{}, and $kT_{bb} < 0.17$\kev{}, respectively. The data/model ratio plots after fitting the continuum with {\tt nthcomp} are shown in Figure~\ref{Fig4}. The fit residuals reveal the presence of several structures. Both spectra show evidence of Fe K$\alpha$ emission in the $5-7$\kev{} energy band, an absorption edge around $\sim 7-8$\kev{}, and a broad Compton hump above $15$\kev{}, indicating the presence of a reflection spectrum. Both spectra show broad, extended to lower energies, and strongly peaked iron emission lines. We thus continued our spectral analysis by including physical reflection models. \\

\begin{figure}
\includegraphics[scale=0.32, angle=-90]{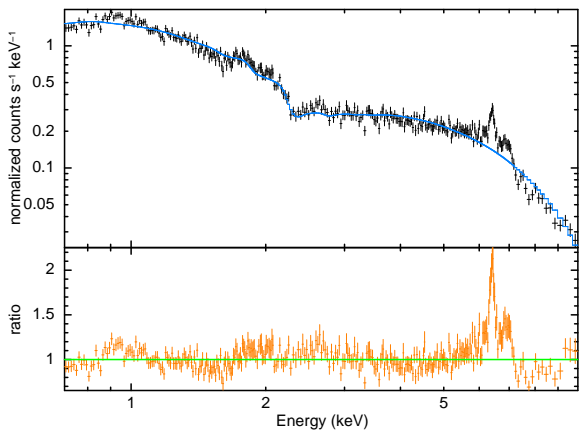}
\caption{The source spectrum was obtained from the \nicer{} observation performed on 22nd March 2019 before the present \nustar{} observation. The lower panel shows the fit residual associated with the continuum model {\tt nthcomp}. The broad emission lines $\sim 6.4$ \kev{}, $\sim 6.9$ \kev{}, and an absorption edge $\sim 7-8$ \kev{} are clearly detected with this \nicer{} spectrum.} 
\label{Fig5}
\end{figure}

In addition, we extracted the spectrum from a \nicer{} observation (\citealt{2016SPIE.9905E..1HG}) which  closely precedes  this \nustar{} observation. This \nicer{} observation (obs ID: 2200400111) was made on 22nd March 2019 prior to this \nustar{} observation. Here also, the continuum is well described by the {\tt nthcomp} component. The broad emission lines $\sim 6.4$\kev{}, $\sim 6.9$\kev{}, and an absorption edge $\sim 7.1$\kev{} are clearly detected with this \nicer{} spectrum (see Figure~\ref{Fig5}). The clear detection of Fe XXVI K$\alpha$ line ($\sim 6.9\kev{}$) indicates the high energy resolution of \nicer{}. We wanted to examine whether the different instruments consistently detect all the residuals, particularly absorption edge $\sim 7-8$\kev{}. However, we did not consider the joint fitting of the \nicer{} and \nustar{} spectra as they are not simultaneous. \\

\subsection{Self-consistent reflection fitting}
When X-rays irradiate the accretion disc, they produce a reflection spectrum including fluorescence lines, recombination, and other emissions \citep{1989MNRAS.238..729F}. It has been observed that in most X-ray sources, the incident emission for the reflection spectrum is generally a hard power-law spectrum. However, in NS systems, the emission from the NS surface/boundary layer may be significant and contribute to the reflection \citep{2008ApJ...674..415C, 2019ApJ...873...99L, 2017ApJ...836..140L}. It is essential to investigate the different possible sources of irradiation as the shape of the reflection continuum depends on the incident spectrum. Since we find a low seed photon temperature of the Comptonization of $< 0.18$\kev{} in our continuum fit with {\tt nthcomp}, we attempt to use {\tt RELXILLCP} to reproduce both Comptonization and reflection spectra correctly. {\tt RELXILLCP} is a relativistic reflection model, a flavor of the {\tt RELXILL} \citep{2014ApJ...782...76G} suite of models that assumes the irradiating continuum is a Comptonization spectrum, treated identically to {\tt nthcomp}. Previously, the model {\tt RELXILLCP} has also been used by other authors to model the reflection component of the spectrum for accreting NSs (\citealt{2019MNRAS.483..767D, 2019ApJ...873...99L, 2022MNRAS.515.3838M}).\\

\begin{figure*}
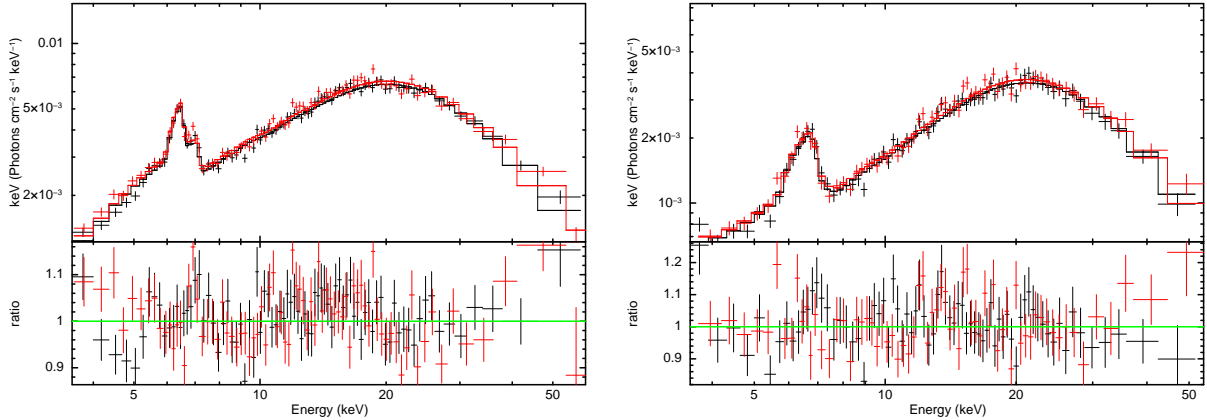

\includegraphics[scale=0.32, angle=-90]{fig7.ps}
\includegraphics[scale=0.32, angle=-90]{fig8.ps}
\caption{Both the spectrum (left: for the high-intensity interval, right: for the low-intensity interval) fitted with the model {\tt const*TBabs*RELXILLCP}. The lower panels of each plot show the ratio of the data to the model in units of $\sigma$. The data were rebinned for better visualization.} 
\label{Fig6}
\end{figure*}

The main parameters of the {\tt RELXILLCP} model are: the inner and outer disc emissivity indices, $q_{1}$ and $q_{2}$, respectively, the inner and outer radii of the disc, $R_{in}$ and $R_{out}$, respectively, the inclination of the system, $i$, the spin parameter, $a$, the power-law photon index, $\Gamma$, the electron temperature of the corona, $kT_{e}$, the ionization parameter, $\xi$, the iron abundance, $A_{Fe}$, the reflection fraction, $r_{refl}$, and the norm which represents the normalization of the model. During the fitting, we choose to have a disc described by a single emissivity index $q_{1}=q_{2}=3$ (a value commonly found in X-ray binaries). The parameters $R_{in}$, $i$, $A_{Fe}$, $\xi$, $\Gamma$, $kT_{e}$ were free to vary but we fixed $R_{out}$ to $1000\;R_{g}$. The dimensionless spin parameter, $a$ for an NS can be approximated as $a\simeq0.47/P_{ms}$ \citep{2000ApJ...531..447B} where $P_{ms}$ is the spin period in ms. However, the burst oscillation frequency of this source is not known yet. So, we fixed $a=0.2$, as for NSs, this value ranges from $0.0$ to $0.3$ and has little effect on the surrounding metric \citep{1998ApJ...509..793M}. Moreover, setting $a$ either $0.0$ or $0.3$ does not yield significant changes in the model parameters. The overall shape of the spectrum is well fitted with this model ({\tt constant*TBabs*RELXILLCP}) with $\chi^2/dof=677/576$ ($\Delta\chi^2=697$ for the $4$ additional parameters) for the high-intensity interval and $\chi^2/dof=299/279$ ($\Delta\chi^2=577$ for the $4$ additional parameters) for the low-intensity interval. The best-fit parameter values have been listed in Table~\ref{parameters0}. The corresponding spectra and the residuals for both spectra are shown in Figure~\ref{Fig6}. \\

\begin{figure*}
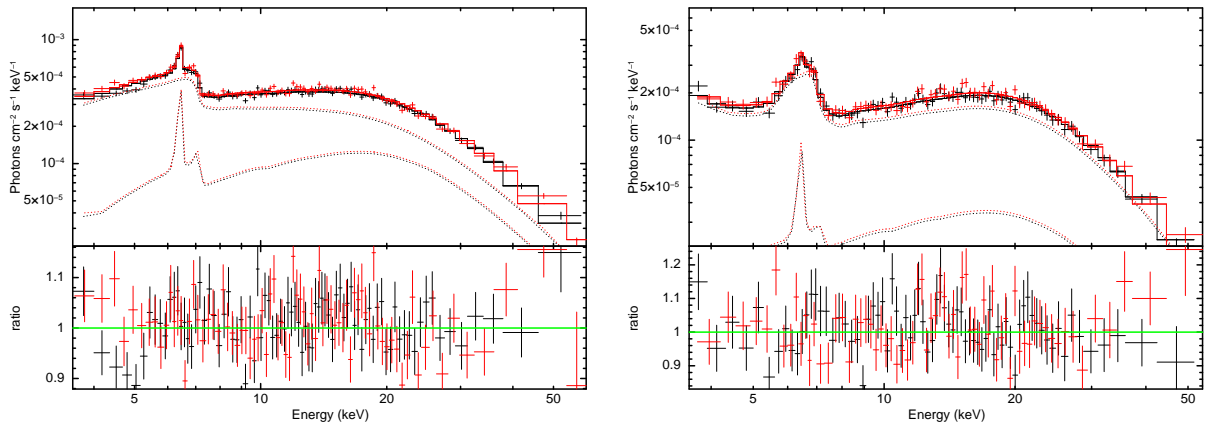

\includegraphics[scale=0.32, angle=-90]{fig9.ps}
\includegraphics[scale=0.32, angle=-90]{fig10.ps}
\caption{Both the spectrum fitted with the model {\tt const*TBabs*(XILLVERCP+RELXILLCP)}. The dotted lines denote the individual model components. The lower panels of each plot show the ratio of the data to the model in units of $\sigma$. The data were rebinned for better visualization.} 
\label{Fig7}
\end{figure*}

Despite the acceptable fit, some residuals still existed near the iron complex $\sim 6.9$ \kev{}, as shown in Figure~\ref{Fig6}. The emission line $\sim 6.9$ \kev{} is also evident from the \nicer{} spectrum of this source. However, this feature in the spectrum is comparatively subtle for the lower-intensity interval. The presence of this emission line suggests that additional model components are required to fit the spectra adequately. We tested two possibilities to understand the nature of this narrow emission line. We initially tried the possibility of a distant reflector, fitting the line with a reflection component {\tt XILLVERCP} \citep{2010ApJ...718..695G}. The {\tt XILLVERCP} model is chosen because its continuum emission model (i.e. {\tt nthcomp}) is the same as that used in the {\tt RELXILLCP} model. We linked all the parameters in {\tt XILLVERCP} to those in {\tt RELXILLCP} as the same source should illuminate both the accretion disc and the distant reflector. We also set the {\tt XILLVERCP} reflection fraction to $-1$, considering that all the emission from this model component is reflected. We allowed the ionization parameter ($\xi$) and the normalization of the {\tt XILLVERCP} component to vary during the spectral fitting. The fit to the spectra with the model {\tt constant*TBabs*(XILLVERCP+RELXILLCP)} was reasonable with $\chi^2/dof=646/574$ ($\Delta\chi^2=31$ for the $2$ additional parameters) for the high-intensity interval spectrum and $\chi^2/dof=285/277$ ($\Delta\chi^2=14$ for the $2$ additional parameters) for the low-intensity interval spectrum. The corresponding spectra and the residuals for both spectra are shown in Figure~\ref{Fig7}. The best-fitting ionization parameters are $2.00_{-0.84}^{+0.09}$ (high-intensity interval) and $\le 1.87$ (low-intensity interval), lower than the one derived from {\tt RELXILLCP} (see Table~\ref{parameters0}). This result does not conflict with our initial assumption of a distant reflector. However, the fit left two important parameters $R_{in}$ and $r_{refl}$ completely unconstrained, and the values were unacceptable. Moreover, freezing $r_{refl}$ to $1$ (a reasonable value for the relativistic reflection model) to continue the fit provides a significantly worse fit for the spectra with $R_{in}$ pegged at its lower value $1\:R_{ISCO}$. We have given the other best-fit model parameter values in Table~\ref{parameters0}, but have not considered this distant reflection scenario further in our discussion.\\

 \begin{table*}
   \centering
\caption{Fit results: Best-fitting spectral parameters of the \nustar{} observations of the source Swift J1858.6-0814 using models  {\tt const*TBabs*RELXILLCP} and {\tt const*TBabs*(RELXILLCP+XILLVERCP)}} 
\begin{tabular}{|p{1.8cm}|p{3.5cm}|p{2.2cm}|p{2.2cm}|p{2.2cm}|p{2.2cm}}
    \hline
     &           & {\tt const*TBabs*RELXILLCP} &  & {\tt const*TBabs*(RELXILLCP+XILLVERCP)} &  \\
    \cline{3-6}
    Component     & Parameter (unit) & High-Intensity Interval & Low-intensity Interval & High-Intensity Interval & Low-Intensity Interval\\
    \hline
    {\scshape Constant} & FPMB (wrt FPMA) & $1.03\pm 0.01$ & $1.04\pm 0.02$ & $1.03\pm 0.01$ & $1.04\pm 0.02$ \\
    {\scshape tbabs}    & $N_{H}$($\times 10^{22}\;\text{cm}^{-2}$) & $13.6_{-0.96}^{+1.07}$  & $5.72_{-0.81}^{+1.29}$ & $11.5_{-2.10}^{+1.15}$ & $\leq 2.94$  \\
    {\scshape relxillCp} & $i$ (degrees) & $21_{-5}^{+6}$  & $44_{-7}^{+2}$ & $36\pm 2$ & $44_{-6}^{+8}$\\
    & $R_{in}$($\times R_{ISCO}$) & $10.5_{-8.8}^{+1.8}$ & $12.4_{-8.7}^{+4.5}$ & $1.0_{peg}^{+2.4}$ & $8.1_{-1.9}^{+2.5}$\\
    & $\rm{log}\:\xi$ (erg cm s$^{-1}$) &  $2.51_{-0.02}^{+0.05}$ &  $2.71_{-0.13}^{+0.02}$ & $2.60_{-0.49}^{+0.11}$ & $2.67_{-0.08}^{+0.15}$ \\
    & $\Gamma$  & $1.58_{-0.03}^{+0.02} $ & $1.32_{-0.02}^{+0.03} $  & $1.53_{-0.03}^{+0.06} $ & $1.40_{-0.02}^{+0.03} $ \\
    & $A_{Fe}$ ($\times \;\text{solar})$   & $0.50_{peg}^{+0.17}$  & $0.50_{peg}^{+0.14}$ & $0.50_{peg}^{+0.16}$ & $0.50_{peg}^{+0.13}$ \\
    & $kT_{e}(\kev)$ &  $15.5\pm 0.88$ & $10.5_{-0.49}^{+1.82}$ & $16.2\pm 0.96$ &  $13.3_{-0.79}^{+0.65}$\\
    & $f_{refl}$   & $2.35_{-1.28}^{+0.99}$ & $Unconstrained$ & $Unconstrained$ & $Unconstrained$ \\
    & norm ($\times 10^{-4}$)   &  $1.74_{-0.29}^{+0.26}$ & $0.37_{-0.16}^{+0.09}$ & $0.77_{-0.08}^{+0.19}$ & $0.01\pm 0.001$ \\
    {\scshape xillverCp} & $\rm{log}\:\xi$ (erg cm s$^{-1}$)  & --  & -- & $2.00_{-0.37}^{+0.09}$ & $\leq 1.87$    \\
    & norm ($\times 10^{-4}$)  & --  & --  & $3.64_{-0.61}^{+1.04}$ & $0.87_{-0.79}^{+0.84}$  \\
    & $f_{refl}$  & -- & -- & $-1$ (freeze) & $-1$ (freeze)  \\
   \hline 
    & $\chi^{2}/dof$ & $677/576$  & $299/279$ & $646/574$ & $285/277$  \\
    \hline
  \end{tabular}\label{parameters0} \\
{\bf Note:} The outer radius of the {\tt RELXILLCP} spectral component was fixed to $1000\;R_{g}$. We fixed emissivity index $q1=q2=3$. The spin parameter ($a$) was fixed to $0.2$ as the fastest spinning NS known in an LMXB has spin $a\sim 0.3$. Other parameters of {\tt XILLVERCP} are linked with the {\tt RELXILLCP}.\\

\end{table*}

\begin{figure*}
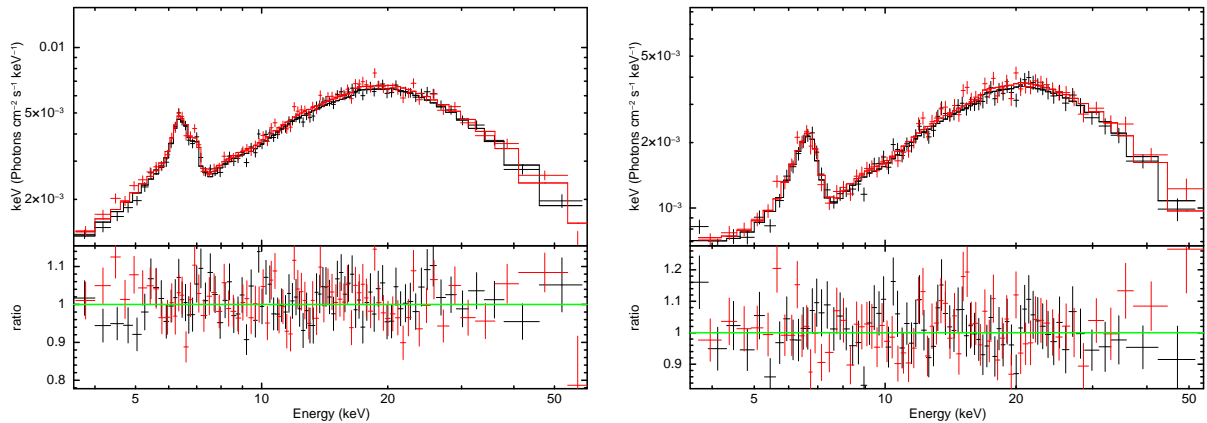

\includegraphics[scale=0.32, angle=-90]{fig11.ps}
\includegraphics[scale=0.32, angle=-90]{fig12.ps}
\caption{Both the spectrum fitted with the model {\tt const*TBabs*ZXIPCF*RELXILLCP}. The lower panels of each plot show the ratio of the data to the model in units of $\sigma$. In both cases, we found that this model perfectly fits the spectra. The data were rebinned for plotting purposes.} 
\label{Fig8}
\end{figure*}

We then tried to examine if the narrow emission line is the re-emission from the disk winds as we observed the sign of an absorption feature ($\sim 7-8 \kev{}$) in the spectra. Outflows in the form of the disc wind have already been detected from this source at both optical and X-ray wavelengths \citep{2020ApJ...893L..19M, 2020MNRAS.498...68B}. Therefore, we included a partially covering absorption model {\tt ZXIPCF} as the detection of such an absorption feature strongly indicates the presence of ionized absorbing plasma in the system. The model was initially developed for AGNs, but later was applied to X-ray binaries as well \citep{2015MNRAS.446.1536P, 2020A&A...635A.209I, 2016MNRAS.461.1917M}. This model considers the case where a fraction $f_{cov}$ of the source is covered by absorbing photoionized matter, with ionization parameter $\xi_{abs}$. Besides these two parameters, the other main parameter of this model is the equivalent hydrogen column, $N_{H}$ of the material. The {\tt ZXIPCF} component is, in principle, able to reproduce also the Fe XXVI line \citep{2022MNRAS.515.3838M}. The fits are substantially improved by the addition of this model, leading to an improvement of $\Delta\chi^2=51$ for three fewer $dof$ in the high-intensity interval spectrum ($\chi^2/dof=626/573$). However, the improvement is comparatively less, $\Delta\chi^2=14$ for three fewer $dof$ for the low-intensity interval spectrum ($\chi^2/dof=285/277$). According to our results, the absorbing material described by {\tt ZXIPCF} covers a fraction of $\sim 80\%$ of the main X-ray source. It is characterized by a medium ionization (log$\xi$ between 1 and 2) and a significant absorbing column density of $N_{H}$ (lower limit of $\sim 5.2\times 10^{23}$ cm$^{-2}$). The best-fit parameters to the model, {\tt constant*TBabs*(ZXIPCF*RELXILLCP)} are given in Table~\ref{parameters1}, and corresponding best-fit spectra with residuals are shown in Figure~\ref{Fig8}. \\

\begin{figure*}
\includegraphics[scale=0.40, angle=0]{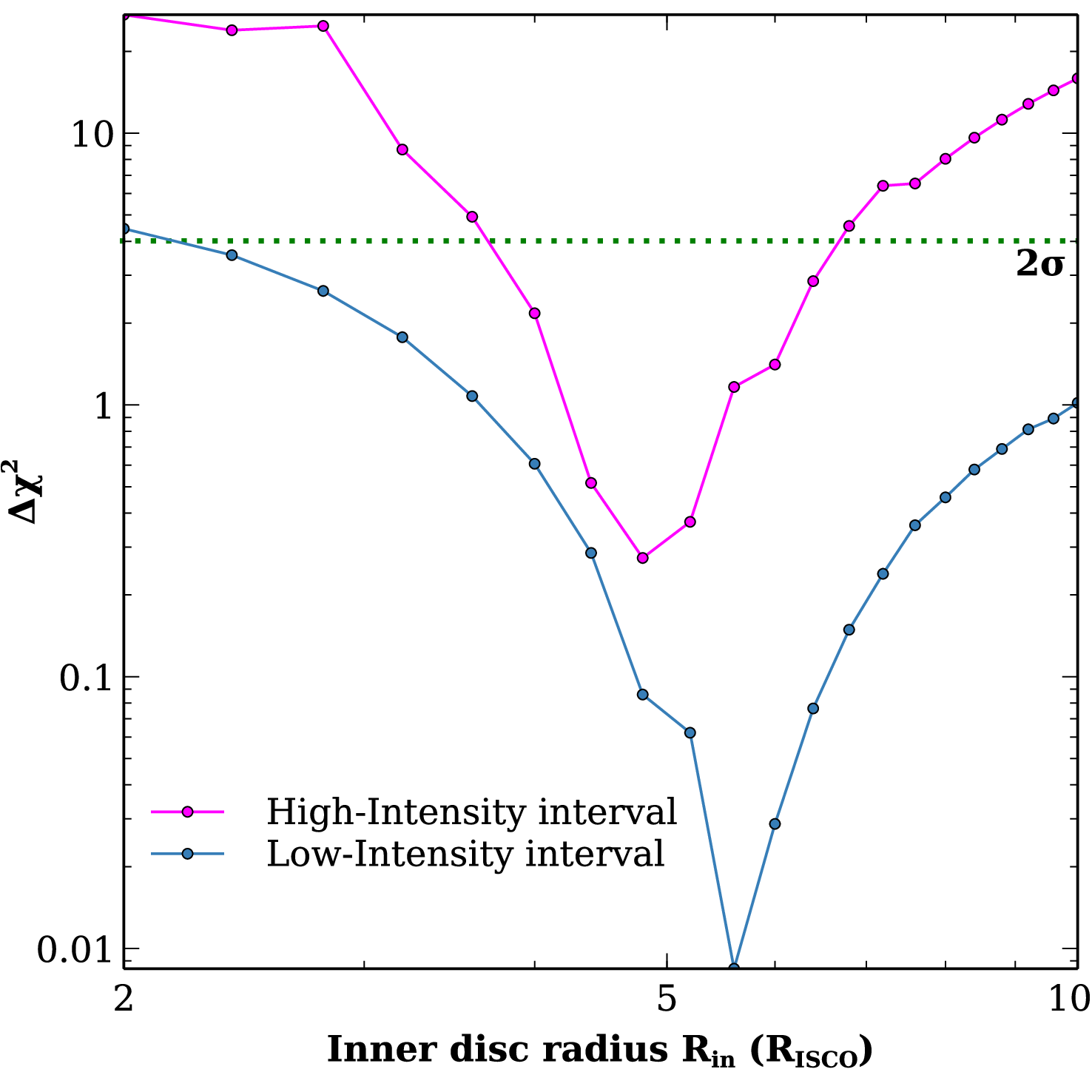}\hspace{2cm}
\includegraphics[scale=0.40, angle=0]{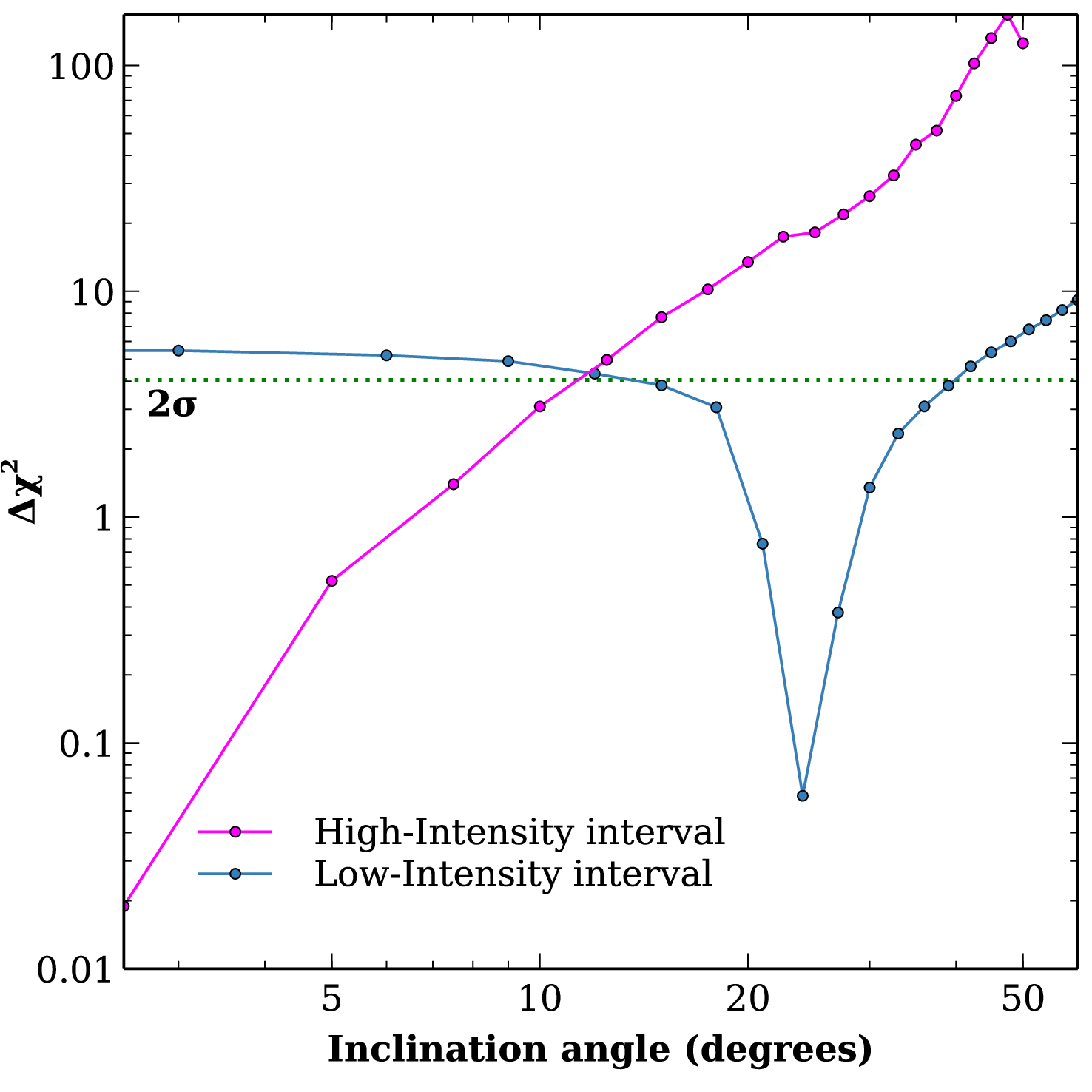}
\caption{The plots show the change in the goodness of fit for the inner disc radius and disc inclination angle. The left panel shows the variation of $\Delta\chi^{2}(=\chi^{2}-\chi_{min}^{2})$ as a function of inner disc radius obtained from the relativistic reflection model ({\tt RELXILLCP}). The right panel shows the variation of $\Delta\chi^{2}(=\chi^{2}-\chi_{min}^{2})$ as a function of the disc inclination angle obtained from the same relativistic reflection model. We varied the disc inclination angle between 0 degrees and 60 degrees..} 
\label{Fig9}
\end{figure*}

From the above, we observed that the overall shape of the complex spectra of this source could be better fitted when we use a partially covering absorption model {\tt ZXIPCF} along with the relativistic reflection model {\tt RELXILLCP}. From the best-fit spectral model, we found an inner disc radius to be $4.87_{-0.96}^{+1.63}\;R_{ISCO}$ (for the high-intensity interval) and $5.68_{-2.78}^{+9.54}\;R_{ISCO}$ (for the low-intensity interval). The inner-disc radius values are consistent with \citet{2020ApJ...890...57H}, implying a significant disc truncation. The inclination is found to be $< 20^{0}$ ($3\sigma$ upper limit) for high-intensity interval and ${25^0}_{-3}^{+8}$ for low-intensity interval. The power-law photon index, $\Gamma$, of the spectra are $1.52_{-0.06}^{+0.04}$ and $1.40_{-0.05}^{+1.52}$ for two intervals, respectively. We found a moderate value $\sim 2.90$ of the disc ionization parameter $\rm{log}\xi$ for both spectra. We further used command {\tt steppar} in {\tt XSPEC} to search the best fit for $R_{in}$ and $i$ for the best-fit model. The left and right panels in Figure~\ref{Fig9} show the $\Delta\chi^2$ of the fit versus the inner disc radius and the disc inclination, respectively, for both the spectrum to our best-fit model {\tt constant*TBabs*(ZXIPCF*RELXILLCP)}. \\

 \begin{table*}
   \centering
\caption{Fit results: Best-fitting spectral parameters of the \nustar{} observations of the source Swift J1858.6-0814 using model:  {\tt const*TBabs$\times$(ZXIPCF$\times$ RELXILLCP)}.} 
\begin{tabular}{|p{1.8cm}|p{4.8cm}|p{2.2cm}|p{2.2cm}}
    \hline
    Component     & Parameter (unit) & High-Intensity Interval & Low-intensity Interval \\
    \hline
    {\scshape Constant} & FPMB (wrt FPMA) & $1.04\pm 0.01$ & $1.04\pm 0.02$ \\
    {\scshape tbabs}    & $N_{H}$($\times 10^{22}\;\text{cm}^{-2}$) &$\leq 11.6$  & $\leq 3.1$   \\
    {\scshape zxipcf} & $N_{H,abs}$($\times 10^{22}\;\text{cm}^{-2}$)  &  $52_{-11}^{+22}$  & $64_{-5}^{+20}$ \\
    & log $\xi_{abs}$  & $1.65_{-0.43}^{+0.29}$  & $1.39_{-0.36}^{+0.30}$   \\
    & $f_{cov}$  & $0.81_{-0.20}^{+0.06}$  & $0.83_{-0.04}^{+0.09}$   \\
    {\scshape relxillCp} & $i$ (degrees) & $< 20^{\dagger}$  & $25_{-3}^{+8}$ \\
    & $R_{in}$($\times R_{ISCO}$) & $4.87_{-0.96}^{+1.63}$ & $5.68_{-2.78}^{+9.54}$\\
    & $\rm{log}\:\xi$(erg cm s$^{-1}$) &  $2.87_{-0.11}^{+0.15}$ &  $2.92_{-0.16}^{+0.50}$\\
    & $\Gamma$  & $1.52_{-0.04}^{+0.06} $ & $1.40_{-0.05}^{+0.23} $ \\
    & $A_{Fe}$ ($\times \;\text{solar})$   & $0.50_{peg}^{+0.18}$  & $1.68_{-0.78}^{+1.52}$\\
    & $kT_{e}(\kev)$ &  $13.5_{-1.70}^{+1.77}$ & $9.25_{-1.08}^{+3.02}$\\
    & $f_{refl}$   & $1.08_{-0.38}^{+0.21}$ & $1$ (frozen)\\
    & norm ($\times 10^{-4}$)   &  $3.77_{-0.33}^{+0.95}$ & $1.97_{-0.18}^{+0.63}$\\
    & $F^{*}_{total}$ ($\times 10^{-10}$ ergs/s/cm$^2$) & $3.72\pm 0.03$ & $1.80\pm 0.02$\\
    & Observed flux ($\times 10^{-10}$ ergs/s/cm$^2$)   & $3.53\pm 0.01$ & $1.64\pm 0.01$ \\
    & $L_{3-79 \kev{}}$ ($\times 10^{36}$ ergs/s) & $9.68 \pm 0.05$ & $4.78\pm 0.04$\\	
  
   \hline 
    & $\chi^{2}/dof$ & $626/573$  & $285/277$ \\
    \hline
  \end{tabular}\label{parameters1} \\
{\bf Note:} The outer radius of the {\tt RELXILLCP} spectral component was fixed to $1000\;R_{g}$. We fixed emissivity index $q1=q2=3$. The spin parameter ($a$) was fixed to $0.2$ as the fastest spinning NS known in an LMXB has spin $a\sim 0.3$. $^{*}$All the unabsorbed fluxes are calculated in the energy band $3-79 \kev{}$ using the {\tt cflux} model component. Luminosity is calculated based upon a distance of $15$ kpc \citep{2020MNRAS.499..793B}. $^{\dagger}$ $3\sigma$ upper limit on the inclination is quoted.\\

\end{table*}

\section{Discussion}
We have analyzed the \nustar{} observation of this source taken on 23rd March 2019. We performed time-resolved spectroscopy on the source by extracting spectra for two different intensity intervals. As stated earlier, the spectra extracted from the high-intensity interval ($> 25$ \ctssec{} during flares) and the low-intensity interval ($<25 \ctssec{}$ during flares) had exposure times of $\sim 19.3$ ks and $\sim 16.7$ ks, respectively. The source was observed with a $3-79 \kev{}$ luminosity of $\sim 9.68\times 10^{36}$ ergs/s and $\sim 4.78\times 10^{36}$ ergs/s for two intensity intervals, respectively assuming a distance of $15$ kpc. The observed luminosity is about $5$ percent of the Eddington luminosity. The luminosity is consistent with the earlier \nustar{} observation made on 3rd November 2018 \citep{2020ApJ...890...57H}. So, the source has continued exhibiting a low luminosity level. The continuum emission of the spectra is well explained by a thermal Comptonization model when soft photons from the accretion disc are considered a source of seed photons for Comptonization. The Comptonization spectrum is due to the cold seed photons ($kT_{bb}< 0.18$ \kev{}), Compton up-scattered by a $\sim 10$\kev{} electron corona into a $\Gamma\sim 1.40$ cut-off power-law. The continuum parameters are consistent with \citet{2020MNRAS.498...68B}. Both spectra show evidence of Fe K$\alpha$ emission in the $5-7$\kev{} energy band, an absorption edge around $7-8$\kev{}, and a broad Compton hump above $15$\kev{}, indicating the presence of a reflection spectrum. We also extracted the \nicer{} spectrum of this source, taken prior to this \nustar{} observation, to explore how the different instruments observe these features. We found that the broad emission lines $\sim 6.4$ \kev{}, $\sim 6.9$ \kev{}, and an absorption edge $\sim 7-8$ \kev{} are clearly detected with the \nicer{} spectrum as well.  \\

We performed a detailed analysis of the resulting reflection spectrum to probe these features further. We found that the observed features described above are well explained by the contribution of two models, one based on relativistic disc reflection ({\tt RELXILLCP})and the other based on partially covering absorption ({\tt ZXIPCF}). The disc reflection is fitted using the relativistic reflection model {\tt RELXILLCP} that assumes the irradiating continuum to be a Comptonization spectrum. Modeling disc reflection with {\tt RELXILLCP}, we obtained that the disc is truncated quite far from the NS surface, at a radius of $\sim(3.9-6.5)R_{ISCO}=(20.8-34.7)R_{g}$ ($43.8-72.8$ km for a $1.4\Msun$ NS) and $\sim(2.9-15.2)R_{ISCO}=(15.5-81.0)R_{g}$ ($32.5-170$ km) respectively for the high-intensity and low-intensity interval (where $R_{ISCO}=5.33R_{g}$ for a spinning NS with $a=0.2$). This truncation radius of the inner accretion disc is consistent with \citet{2020ApJ...890...57H}. In the disc truncation scenario, the position of the inner disc radius should move towards the NS surface with increasing mass accretion rate \citep{2016ApJ...831...45C}. This behavior is consistent in this case though there is larger uncertainty in the $R_{in}$ during the low-intensity interval. The reflection fit yielded a low inclination estimate of $< 20^{0}$ for the high-intensity interval and $(22^{0}-33^{0})$ for the low-intensity interval. This inclination suggests that the disc is viewed almost face-on. A low inclination is consistent with previous spectral results obtained with the \nustar{} spectrum \citep{2020ApJ...890...57H}. The reflection spectrum also revealed that the accretion disc is highly ionized, log$\xi$ is consistent with $\sim 3$, and the iron abundance is comparable to the solar abundance.   \\

As the disc reflection model shows evidence of a significant disc truncation, we examined different possibilities responsible for the same. We first tested whether the boundary layer is responsible for disc truncation by calculating its maximum radial extension from the NS surface. To do so, we first estimated the mass accretion rate per unit area, using Equation (2) of \citet{2008ApJS..179..360G}
\begin{equation}
\begin{split}
\dot{m}=&\:6.7\times 10^{3}\left(\frac{F_{p}\:c_\text{bol}}{10^{-9} \text{erg}\: \text{cm}^{-2}\: \text{s}^{-1}}\right) \left(\frac{d}{10 \:\text{kpc}}\right)^{2} \left(\frac{M_\text{NS}}{1.4 M_{\odot}}\right)^{-1}\\
 &\times\left(\frac{1+z}{1.31}\right) \left(\frac{R_\text{NS}}{10\:\text{km}}\right)^{-1} \text{g}\: \text{cm}^{-2}\: \text{s}^{-1}.
 \end{split} 
\end{equation}
The above equation outturns a mass accretion rate of $1.6\times 10^{-9}\;M_{\odot}\;\text{y}^{-1}$ for the high-intensity interval and $0.76\times 10^{-9}\;M_{\odot}\;\text{y}^{-1}$ for the low-intensity interval at a persistent flux $F_{p}=3.7\times 10^{-10}$ erg~s$^{-1}$ cm$^{-2}$ and $F_{p}=1.8\times 10^{-10}$ erg~s$^{-1}$ cm$^{-2}$, assuming the bolometric correction $c_{bol} \sim 1.38$ for the nonpulsing sources \citep{2008ApJS..179..360G}. Here we assume $1+z=1.31$ (where $z$ is the surface redshift) for an NS with mass ($M_{NS}$) 1.4 $M_{\odot}$ and radius ($R_{NS}$) $10$ km. We used this mass accretion rate to estimate the maximum radial extension of the boundary layer (from Equation (2) of \citet{2001ApJ...547..355P}. We found the maximum value of the boundary layer to extend to $R_{BL}\sim 5.7\;R_{g}$ for the high-intensity interval and $R_{BL}\sim 5.4\;R_{g}$ for the low-intensity interval (assuming $M_{NS}=1.4\:M_{\odot}$ and $R_{NS}=10$ km). The actual value may be larger than this if we account for the viscous effects and the spin of this layer. Still, the extent of the boundary layer regions  is too small to account for the disc position. Thus, it is implausible that the boundary layer is responsible for the disc truncation. \\

It has been observed that in the case of many NS LMXBs, the accretion disc is truncated at moderate radii due to the pressure exerted by the magnetic field of the NS. One can estimate the magnetic field strength if it is truncated at the magnetospheric radius. An upper limit of the magnetic field strength of the NS can be estimated with the inferred inner disc radius from the \nustar{} fits. Equation (1) of \citet{2009ApJ...694L..21C} gives the following expression for the magnetic dipole moment,
\begin{equation}
\begin{split}
\mu=&3.5\times 10^{23}k_{A}^{-7/4} x^{7/4} \left(\frac{M}{1.4 M_{\odot}}\right)^{2}\\
 &\times\left(\frac{f_{ang}}{\eta}\frac{F_{bol}}{10^{-9} \text{erg}\: \text{cm}^{-2}\: \text{s}^{-1}}\right)^{1/2}
 \frac{D}{3.5\: \text{kpc}} \text{G}\; \text{cm}^{3},
\end{split} 
\end{equation}
where $\eta$ is the accretion efficiency in the Schwarzschild metric, $f_{ang}$ is the anisotropy correction factor (close to unity) and $k_{A}$ is a geometrical coefficient expected to be $\sim(0.5-1.1)$. We note that \citet{2009ApJ...694L..21C} modified $R_{in}$ as $R_{in}=x\:GM/c^{2}$ introducing a scale factor $x$. We assumed $k_{A}=1$, $f_{ang}=1$ and $\eta=0.2$ (as reported in \citealt{2009ApJ...694L..21C} and \citealt{2000AstL...26..699S}) and used $0.1 - 100$\kev{} flux as the bolometric flux ($F_{bol}$) of $\sim 3.8\times 10^{-10}$ erg~s$^{-1}$ cm$^{-2}$. Using the $R_{in}$ value found in low-intensity interval ($\leq 81\;R_{g}$), we obtained $\mu \leq 4.6\times 10^{27}$ G cm$^{3}$, where the upper bound of $R_{in}$ considers both the high and the low-intensity intervals. This leads to an upper limit of the magnetic field strength of $B\leq 9.2\times 10^{9}$ G at the magnetic poles (for an NS mass of $1.4\:M_{\odot}$, a radius of $10$ km, and a distance of $15$ kpc). A relatively low magnetic field strength was also predicted by \citet{2020MNRAS.499..793B}.  \\

Suppose disc truncation is a consequence of the NS dipolar magnetic field halting the accretion flow. In that case, the inner edge of the disc obtained from the reflection fit must coincide with such magnetospheric radius ($R_{M}$). At this radius, the magnetospheric pressure equals the ram pressure of the accreted matter. The accretion disc is interrupted at the magnetospheric radius for disc-accretion, given by \citep{1979ApJ...232..259G}
\begin{equation}
R_{M}=1300\:L_{37}^{-2/7}\:M^{1/7}\:R_{6}^{10/7}\:B_{12}^{4/7} \;\text{km},
\end{equation}
where $M$ is the mass of NS in $1.4\:M_{\odot}$ units, $R_{6}$ is the radius in units of $10^{6}$ cm, $B_{12}$ is the surface magnetic field strength in $10^{12}$ G units, and $L_{37}$ is the accretion luminosity in units of $10^{37}$ erg~s$^{-1}$. Taking the upper limit of the $B$ field and the luminosity (estimated earlier), we estimated the magnetospheric radius $\sim 47$ km for the high-intensity interval and $\sim 71$ km for the low-intensity interval from the central object (assuming $M_{NS}=1.4\:M_{\odot}$ and $R_{NS}=10$ km).
This is in good agreement with the position of $R_{in}$ for both the spectrum. This strengthens the possibility that the magnetosphere truncates the disk, although various other possible explanations exist for a truncated disk \citep{1997ApJ...489..865E}. For convenience, we have listed all the estimated values of the parameters in Table~\ref{parameters2}.\\

 \begin{table}
   \centering
\caption{Estimation of various parameters.} 
\begin{tabular}{|p{2.5cm}|p{2.0cm}|p{2.0cm}}
\hline
Parameters (Unit) & High-Intensity Interval & Low-intensity Interval \\  
\hline
$R_{in}$ ($ R_{g}$) &   $20.8-34.7$ &  $15.5-81.0$  \\

$\dot{m}$ ($10^{-9}\;M_{\odot}\;\text{y}^{-1}$) & $\sim 1.6$ & $\sim 0.76$ \\

$R_{BL, max} (R_{g})$ & $\sim 5.7$ & $\sim 5.4$ \\

$B_{max}$ ($ 10^{9}$ G) &  $\le 2.1$ &  $\le 9.2$\\

$R_{M}$ (km)  & $\sim 47$  &  $\sim 71$ \\
\hline
\end{tabular}\label{parameters2} \\
See text (mainly discussion section) for the details of the calculations of the parameters.
\end{table}

Our best-fit spectral model revealed a very large value of the neutral hydrogen (HI) absorbing column density for both the spectrum (i.e., $\sim 11\times 10^{22}$ cm$^{-2}$ for high-intensity interval and $\sim 3\times 10^{22}$ cm$^{-2}$ for low-intensity interval). A very large absorbing column density for this source was also reported by \citet{2020ApJ...890...57H}. However, we found that the source observed fluxes do not differ much from its unabsorbed fluxes (i.e., intrinsic fluxes) in both intensity intervals (see Table~\ref{parameters1}). A large absorbing column density should cause a noticeable change in the observed source's flux. The absence of such behavior indicates that the slight change in the source's flux does not account for this significant absorption. Moreover, we note that the total Galactic HI absorbing column density in this direction is only $N_{H}\sim 1.8\times 10^{22}$ cm$^{-2}$ \citep{1990ARA&A..28..215D}. Therefore, it supports that this absorption must be intrinsic to the binary itself (see \citet{2020ApJ...890...57H} for further discussion). In our spectral fitting, we found that the fit of the energy spectrum is adequately improved if a partially covering absorber is added to the reflection model. According to our results, the absorbing material described by the {\tt ZXIPCF} covers a fraction of $\sim 80\%$ of the main X-ray source, characterized by a medium ionization (log$\xi$ between 1 and 2). We further found that the partially covering absorber has a significant absorbing column density of $N_{H,abs}=52_{-11}^{+22} \times 10^{22}$ cm$^{-2}$ (for the high-intensity interval). According to the high covering fraction, this absorbing material could be interpreted as an ionized thermal plasma coating the optically thin disc and seen face-on, most likely associated with the Fe XXVI wind. The source properties are consistent with those observed by \citet{2018ATel12220....1R}, suggesting a significant intrinsic absorption in the source during this flaring outburst state. However, the actual geometry of the absorber may be different to cover the compact X-ray emission region partially. Our spectral analysis also shows a slight hardening and increase of the absorbing column density ($N_{H,abs}=64_{-5}^{+20} \times 10^{22}$ cm$^{-2}$) between the high-intensity interval and low-intensity interval where extensive flaring activity is mostly absent. It causes a reduction in the observed luminosity as the highly dense local ionized absorber could also produce significant Thomson scattering. The increase in the absorption might reflect an increase in the mass of the ionized thermal plasma associated with the optically thin disc. This behavior of this source is very much similar to the other NS LMXB IGR J17407-2808, which also shows strong flaring activity in the X-ray light curve \citep{2023arXiv230408816D}. Recently, Relativistic X-ray reflection and photoionized absorption have been detected in bright NS LMXB GX 13+1, where significant flaring activity has also been observed in the X-ray light curve \citep{2023MNRAS.522.3367S}.  \\

Various characteristics like large-scale flaring activity in the X-ray through optical bands, large intrinsic variable absorbing column density, lack of a well-defined spectral state, and disc winds detected in optical bands are very similar to the well-known BH transient sources V404 Cyg and V4641 Sgr. The detection of varying P-Cygni profiles in the optical spectra of this source suggests that it has a high velocity ($\sim 25000$ km s$^{-1}$) outer accretion disc winds, similar to those observed from both the sources V404 Cyg and V4641 Sgr \citep{2019ATel12881....1M, 2002A&A...385..904R}. The orbital inclination angle for both V404 Cyg and V4641 Sgr are relatively large, $\sim 67^{0}$ and $\sim 72^{0}$, respectively \citep{2010ApJ...716.1105K, 2014ApJ...784....2M}. A large misalignments (possibly $\sim 30^{0}$ - $50^{0}$) between the orbit of the system and the inner accretion disc have been observed for these two sources \citep{2002MNRAS.336.1371M, 2019Natur.569..374M}. Here for the source Swift~J1858.6-0814, we observed an upper limit of the disc inclination angle $< 33^{0}$ for the low-intensity interval. The lower limit of the inclination of the binary orbit for this source was found to be $\sim 70^{0}$ by \citet{2021MNRAS.503.5600B}. If the disc inclination angle is compared to the binary inclination, then a large misalignment of $\sim 40^{0}$ is observed which is similar to the sources V404 Cyg and V4641 Sgr. \\

Similar mismatches in inclination between the orbit of the system and the inner accretion disc have previously been observed for other BH binary systems as well \citep{2019MNRAS.490.1350B, 2019ApJ...882..179C, 2020MNRAS.493.5389F}. This discrepancy is attributed to the pronounced inner disk warp, so the inner disc is at a different inclination to the binary orbit \citep{2019ApJ...882..179C}. The warped inner disc has been connected to the rapid rotation of a misaligned BH. However, the difference in inclination of $\sim 40^{0}$ would be an unusually strong wrap for the accreting NS systems where the binary mass ratio $q$ is typically observed $\lesssim 1$ ($q=M_{2}/M_{1}$ where $M_{1}$ and $M_{2}$ are the mass of the compact and companion object, respectively). The $q$ value for the NS LMXB is much larger than for BH X-ray binaries, typically ranging $\sim 0.05 - 0.1$ \citep{2019ApJ...882L..21T}. In our case, one of the possible reasons for this mismatch could be the obscuration of the inner regions of the accretion disc by a flared disc as indicated by \citet{2020ApJ...890...57H} and \citet{2021MNRAS.503.5600B}. Any vertical structure of the inner disk (possibly with clumps of material) may obscure blueward line emission, leading to an inferred disk inclination much lower than the actual value in reflection modeling \citep{2018ApJ...855..120T}. Besides these, more complex irradiating source geometries (e.g., vertically or horizontally extended corona, outflowing relativistic jet) and high density of the disc could contribute to the inclination estimates. An insight into the proper measurement of disk inclination may be obtained by a detailed study of relativistic reflection modeling, which is beyond the scope of the present work. \\

While the sources V404 Cyg and V4641 Sgr are similar to Swift~J1858.6-0814 in many ways, one crucial difference still exists between the first two systems and the current source. Both the sources V404 Cyg and V4641 Sgr are much luminous compared to Swift~J1858.6-0814 as the formerly mentioned sources are accreting at Eddington limit, whereas the  source under consideration was accreting at only a few percent ($\sim 5\%$) of the Eddington luminosity. The less amount of observed luminosity can possibly be explained by large enough absorbing column densities ($N_{H}> 10^{23}$ cm$^{-2}$) for which scattering processes become important. Finally, for further studies of different source properties, including its X-ray variability and the spectral profiles associated with the iron (in both emission and absorption) in great detail, one needs multi-wavelength observations.

\section{Data availability}
This research has made use of data obtained from the HEASARC, provided by NASA's Goddard Space Flight Center. Both observational data sets with Obs. IDs $90501309002$ (\nustar{}) and $2200400111$ (\nicer{}) dated 23rd March 2019 and 22nd March 2019, respectively are in public domain put by NASA at their website https://heasarc.gsfc.nasa.gov.  
 
\section{Acknowledgements}
 We thank the anonymous referee for his/her valuable comments, which helped to improve this manuscript considerably. This research has made use of data and/or software provided by the High Energy Astrophysics Science Archive Research Centre (HEASARC). This research also has made use of the \nustar{} data analysis software ({\tt NuSTARDAS}) jointly developed by the ASI Space Science Data Center (SSDC, Italy) and the California Institute of Technology (Caltech, USA). This work is supported by NASA through the NICER mission. ASM and BR would like to thank Inter-University Centre for Astronomy and Astrophysics (IUCAA) for their facilities extended to him under their Visiting Associate Programme.

\def\apj{ApJ}
\def\apjl{ApJl}
\def\pasp{PASP} \def\mnras{MNRAS} \def\aap{A\&A} \def\physerp{PhR} \def\apjs{ApJS} \def\pasa{PASA}
\def\pasj{PASJ} \def\nat{Nature} \def\memsai{MmSAI} \def\araa{ARAA} \def\iaucirc{IAUC} \def\aj{AJ} \def\aaps{A\&AS}
\bibliographystyle{mn2e}
\bibliography{aditya}

\end{document}